\documentclass[%
 reprint,
 showpacs,preprintnumbers,
 amsmath,amssymb,
 aps,
]{revtex4-1}

\usepackage{graphicx}
\usepackage{dcolumn}
\usepackage{bm}
\usepackage{braket}
\usepackage[version=3]{mhchem}

\newcommand{\figref}[1]{Fig.~\ref{#1}}

\newcommand{\mrm}{\mathrm}

\newcommand{\mcl}{\mathcal}

\newcommand{\lag}{\langle}
\newcommand{\rag}{\rangle}
\newcommand{\br}[1]{\left( #1 \right)}

\newcommand{\ev}[1]{\left\langle #1 \right\rangle}
\newcommand{\av}[1]{\left| #1 \right|}
\newcommand{\nrm}[1]{\left\| #1 \right\|}

\newcommand{\dr}{\mrm{d}}

\newcommand{\alp}{\alpha}

\newcommand{\tht}{\theta}

\newcommand{\ham}{\mcl{H}}

\newcommand{\pri}{\prime}

\newcommand{\reff}{\mrm{eff}}

\newcommand{\rAA}{\mrm{\AA}}
\usepackage{color,ulem}

\begin{document}

\title{Magnetic orders induced by RKKY interaction \\in Tsai-type quasicrystalline approximant Au-Al-Gd}

\author{Haruka Miyazaki}
\affiliation{Department of Applied Physics, Tokyo University of Science, Katsushika, Tokyo 125-8585, Japan}
\author{Takanori Sugimoto}%
\email{sugimoto.takanori@rs.tus.ac.jp}
\affiliation{Department of Applied Physics, Tokyo University of Science, Katsushika, Tokyo 125-8585, Japan}
\author{Katsuhiro Morita}
\affiliation{Department of Applied Physics, Tokyo University of Science, Katsushika, Tokyo 125-8585, Japan}
\author{Takami Tohyama}
\affiliation{Department of Applied Physics, Tokyo University of Science, Katsushika, Tokyo 125-8585, Japan}

\date{\today}

\begin{abstract}
Recent experimental study on Tsai-type quasicrystalline approximant Au-Al-Gd has revealed the presence of magnetic orders and phase transitions with changing the Au/Al concentration. Motivated by the experiment, we theoretically investigate whether a successive change of magnetic orders occurs in a minimal magnetic model including the RKKY interaction only. We find that the model induces multifarious magnetic orders depending on the Fermi wavenumber and gives a good starting point for understanding the experimental observation.  In addition, we predict the presence of an undiscovered novel magnetic order called cuboc order at large Fermi wavenumber region.
\end{abstract}

\pacs{Valid PACS appear here}
\maketitle

Recent experimental studies on Tsai-type quasicrystals including rare-earth ions~\cite{Tsai94,Tsai00} have successively provided startling discoveries of novel phenomena: valence fluctuation~\cite{Kawana10,Watanuki12}, quantum criticality~\cite{Deguchi12,Alves16} and superconductivity~\cite{Kamiya18,Deguchi15}. These phenomena are induced by strongly-correlated electrons originating from rare-earth ions, particularly Yb and Ce. The physics behind the phenomena might be similar to that of heavy-fermion materials. However, crucial roles of quasi-periodicity in these phenomena still remains an open question in spit of recent theoretical efforts~\cite{Watanabe10,Watanabe12,Watanabe17,Sakai17,Sakai19}. 

Tsai-type quasicrystalline approximants, which have the same local structure as the quasicrystals but keep the translational symmetry~\cite{Tsai13}, have also attracted much attentions due to experimental discovery of various magnetic orders~\cite{Goldman14}, e.g., ferromagnetism, antiferromagnetism, and spin-glass-like magnetism, both in binary~\cite{Tamura10,Kim12,Tamura12,Mori12,Das17} and ternary~\cite{Hiroto13,Hiroto14,Gebresenbut14,Ishikawa16,Gebresenbut16,Ishikawa18,Sato19,Yoshida19} compounds. This is in contrast to quasicrystal where there is no report on magnetic order so far~\cite{Fukamichi86,Warren86,Hattori95,Charrier97,Noakes98,Islam98,Sato98,Sato00,Scheffer00}. In the approximants, the magnetic moments located on rare-earth ions can interact each other via the Rudderman-Kittel-Kasuya-Yosida (RKKY) interaction~\cite{Ruderman54,Kasuya56,Yosida57}. Atomic composition in the approximants is a controlling parameter of the Fermi wavenumber $k_F$, which changes spacial extension of the RKKY interaction. Therefore, an idea that the RKKY interaction as a function of $k_F$ is crucial for understanding the various magnetic orders has been put forward~\cite{Goldman14}. However, the crystal structure of Tsai-type quasicrystals/approximants is too complicated to theoretically analyze the magnetic behaviors. Actually, magnetic orders in three dimensional quasicrystals/approximants remain unclear from theoretical viewpoints, contrary to pioneering works on magnetism in low-dimensional quasiperiodic systems~\cite{Achiam86,Godreche86,Okabe88,Ledue93,Vedmedenko03,Lifshitz97,Wessel03,Jagannathan05,Thiem15}. 

Au-Al-Gd system is one of 1/1 Tsai-type quasicrystalline approximants Au-X-RE (X=Al, Si; RE=Gd, Tb, Ho, Dy) showing magnetic orders~\cite{Hiroto14,Ishikawa18,Sugimoto16}. With increasing Au concentration, i.e., decreasing $k_F$, magnetism in Au-Al-Gd changes from spin glass to ferromagnetism and to antiferromagnetism~\cite{Ishikawa16,Ishikawa18}. Since single-ion anisotropy due to spin-orbit coupling and crystal field is weak on a Gd ion, the Au-Al-Gd system is a good material to investigate interplay of RKKY interactions and the complicated structure in the Tsai-type approximants.

In this paper, we theoretically investigate magnetic states in Au-Al-Gd based on the classical approximation of localized quantum spins on Gd ions. The approximation is justified by the fact that (i) magnetic moment on Gd ions has been estimated to be $\mu_\reff=8.74\mu_B$ from magnetic susceptibility~\cite{Ishikawa16,Ishikawa18} (cf. the effective moment of a single Gd$^{3+}$ ion is $7.94\mu_B$), which is large enough to be approximated as a classical spin, and (ii) magnetic properties in a Tsai-type 1/1 approximant Au-Si-Tb with the same crystal structure as Au-Al-Gd but with strong single-ion anisotropy~\cite{Das17,Sato19} has been described well by the classical approximation~\cite{Sugimoto16}. Calculating possible magnetic orders in a simple model with both Gd ions in the Tsai-type 1/1 approximant structure and RKKY interaction, we confirm a good qualitative accordance with the experimental change of magnetic orders~\cite{Ishikawa16,Ishikawa18}. In addition, we predict an undiscovered magnetic order called cuboc order at large $k_F$ region. 

\begin{figure}[tbp]
 \begin{center}
  \includegraphics[width=0.4\textwidth]{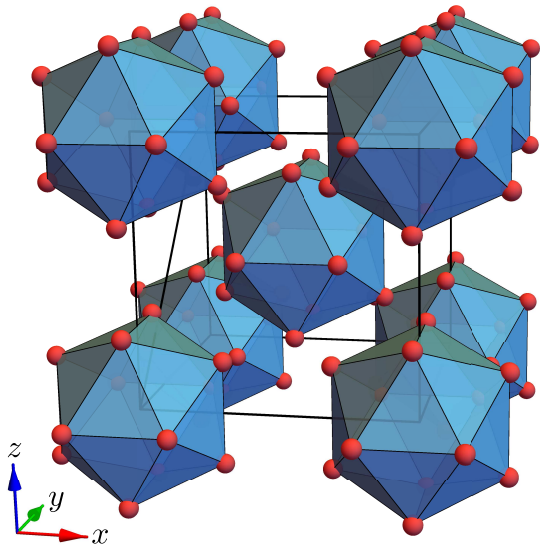}
 \end{center}
 \caption{Magnetic lattice of the Au-Al-Gd alloy. Red balls denote Gd ions on vertices of icosahedron. The icosahedra compose bcc structure. 
 }
 \label{fig1}
\end{figure}

A unit cell in the 1/1 approximant includes two Tsai clusters.  Gd ions occupy icosahedral vertices of the Tsai cluster. The lattice of Gd ions corresponds to body-centered cubic (bcc) of icosahedrons (see \figref{fig1}), and the localized magnetic moments are located on the Gd ions. Thus, there are 24 spins ($n_s=24$) in the unit cell. 

The minimum model Hamiltonian with the RKKY interaction only is given by
\begin{equation}
  \ham = -\sum_{\av{\bm{r}-\bm{r}^\pri}<R_c} J_{\av{\bm{r}-\bm{r}^\pri}} \bm{S}_{\bm{r}}\cdot\bm{S}_{\bm{r}^\pri}, \label{ham}
\end{equation}
where the exchange energy between two spins, $\bm{S}_{\bm{r}}$ at $\bm{r}$ and $\bm{S}_{\bm{r}^\pri}$ at $\bm{r}^\pri$, is given by $J_{\av{\bm{r}-\bm{r}^\pri}}=J f(2k_F |\bm{r}-\bm{r}^\pri|)$ ($J>0$) with the function of Friedel oscillation $f(x)=(-x\cos x +\sin x)/x^4$. In this work, we use a classical approximation of quantum spin and thus $\bm{S}_{\bm{r}}$ is expressed as a three-dimensional normalized vector with $\av{\bm{S}_{\bm{r}}}=1$. In the 1/1 approximant Au-Al-Gd, the magnetic phases successively change with changing the Au concentration~\cite{Ishikawa16,Ishikawa18}. Here we assume that $k_F$ is determined by the electron density $n$ associated with the Au concentration via $k_F=(3\pi^2n)^{1/3}$. In the Hamiltonian \eqref{ham}, we introduce a cutoff range $R_c$ of the RKKY interaction for simplicity of calculation. 

We perform numerical calculation with the classical Monte-Carlo (MC) method to obtain spin configurations at zero temperature. In this calculation, we use single-update heat-bath method combined with the over-relaxation technique and the temperature-exchange method. The system size is set to $N_c=8\times 8\times 8$ unit cells, corresponding to $N_s=N_cn_s=12288$ spins, with periodic boundary condition. We take the number of replicas $N_R=200$, the number of MC steps for relaxation $N_{\mrm{MC}}=2400$, and the lowest temperature $T_M/J=1.0\time 10^{-7}$. After performing the MC simulation, we update the state until the energy converges at $T=0$ to obtain the ground state. We use $R_c=50 \rAA$, which is larger than three times as long as the lattice unit $a=14.7 \rAA$~\cite{note}. $k_F$ is changed from $1.28 \rAA^{-1}$ to $1.61 \rAA^{-1}$, where the experimentally determined $k_F$ in Au-Al-Gd is included~\cite{Ishikawa16,Ishikawa18}.  

\begin{figure}[tbp]
 \begin{center}
  \includegraphics[width=0.49\textwidth]{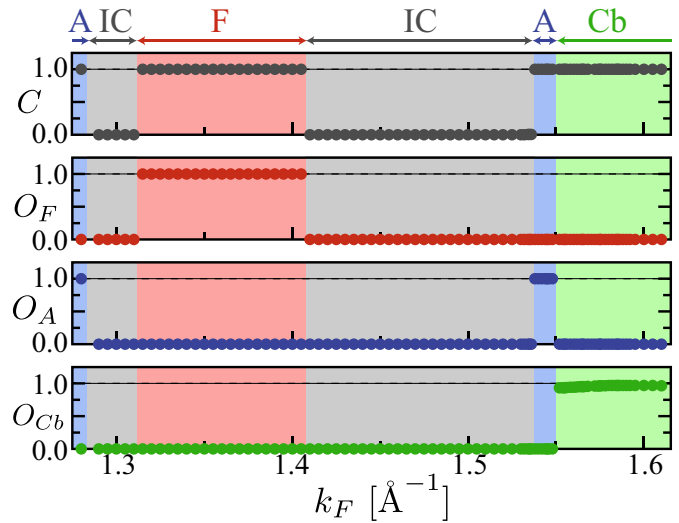}
 \end{center}
 \caption{Phase diagram of the Hamiltonian \eqref{ham} at zero temperature. Four panels represent commesurability $C$, ferromagnetic order $O_F$, antiferromagnetic order $O_A$, and cuboc order $O_{Cb}$, as a function of the Fermi wavenumber $k_F$. Color-shaded regions denote commensurate phases, where only one of the order parameters $O_F$, $O_A$, $O_{Cb}$ is finite: the ferromagnetc (F) phase in red region, the antiferromagnetc (A) phase in blue region, the cuboc (Cb) phase in green region. The gray region corresponds to the incommensurate (IC) phase. 
 }
 \label{fig2}
\end{figure}

To classify the spin configurations of the ground state using our method, we consider commensurability defined by
\begin{equation}
  C=\frac{1}{n_s}\sum_i \nrm{\ev{\bm{S}_{i}}}
\end{equation}
with the averaged magnitude of spins $\ev{\bm{S}_{i}}$ over all unit cells: $\ev{\bm{S}_{i}}=N_c^{-1}\sum_j \bm{S}_{i,j}$, where $\bm{S}_{i,j}$ represents the $i$-th spin in the $j$-th unit cell. If the spin configuration is invariant with respect to translation of the lattice, the commensurability equals to unity. It should be noted that the commensurability is zero in the case of two sublattice configuration; e.g., a spin in a unit cell is directed opposite to the corresponding spin in neighboring unit cells. Thus, this quantity is a measure of ferroic character in the spin configuration. The top panel in \figref{fig2} shows the commensulabirity $C$ as a function of $k_F$. The value of $C$ alternates between 1 and 0 from $k_F=1.28 \rAA$ to $1.61 \rAA$, that is, magnetic state switches from commensurate to incommensurate states and vise versa several times in this region.

To clarify the commensurate state in detail, we consider ferromagnetic and antiferromagnetic order parameters defined by
\begin{align}
  O_F&=\frac{1}{n_s}\nrm{\sum_i \ev{\bm{S}_{i}}}, \\
  O_A&=\frac{1}{n_s}\nrm{\br{\sum_{i\in\mrm{CCI}} \ev{\bm{S}_{i}}-\sum_{i\in\mrm{BCI}} \ev{\bm{S}_{i}}}},
\end{align}
respectively, where CCI (BCI) represents the set of positions on the cubic-cornered icosahedron (body-centered icasahedron) in the unit cell. The second and third panels in \figref{fig2} show these order parameters. We find that these order parameters are equal to 0 or 1 in the whole region. $O_F=1$ represents perfect ferromagnetism, while $O_A=1$ represents an antiferromagnetic phase with N\'eel order where spins in an icosahedron exhibit ferromagnetc order while spins located in a neighboring icosahedron have the opposite direction. Antiferromagnetic phases are located in narrow regions in \figref{fig2}. 

\begin{figure}[tbp]
 \begin{center}
  \includegraphics[width=0.51\textwidth]{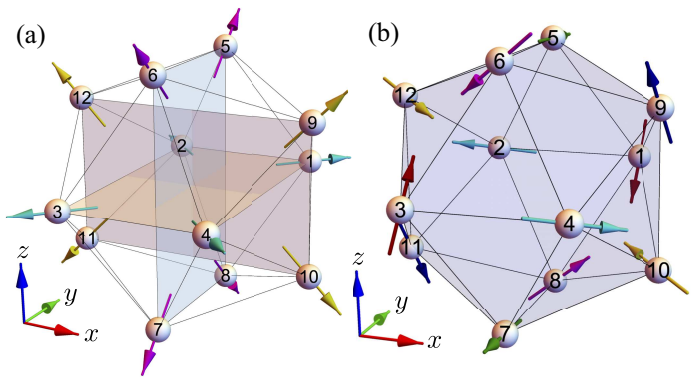}
 \end{center}
 \caption{Cuboc order. (a) An example of spin configuration with $O_{Cb}=1$. Each plane has four spins denoted by the same color, e.g., the first to fourth spins belong to the $xy$ plane. Neighboring spins in the same plane exhibit 90 degree order. (b) Spin configuration observed in the MC calculation with $k_F=1.60\ \rAA^{-1}$. The spin configurations (a) and (b) are equivalent under a global O(3) rotation. 
 }
 \label{fig3}
\end{figure}

In \figref{fig2}, there is a commensurate region of neither ferromagnetic nor antiferromagnetic states above $k_F=1.55  \rAA^{-1}$. In the bottom panel of \figref{fig2}, we successfully identify this magnetic phase with so-called {\it cuboc} order using a corresponding order parameter~\cite{Domenge05,Domenge08,Messio11} defined by
\begin{equation}
  O_{Cb}=\av{\ev{\bm{K}_{xy}}\cdot\ev{\bm{K}_{yz}}\times\ev{\bm{K}_{zx}}}
\end{equation}
with the averaged vector chirality $\ev{\bm{K}_\alp} $ of neighboring spins in the $\alp=xy, xz$, and $zx$ planes given by
\begin{equation}
  \ev{\bm{K}_\alp} = \frac{1}{8}\sum_{\lag i,i^\pri\rag\in \alp\, \mrm{plane}} \ev{\bm{S}_i}\times\ev{\bm{S}_{i^\pri}},
\label{6}
\end{equation}
where the summation runs over all neighboring bonds $\lag i,i^\pri\rag$ in the $\alpha$ plane of icosahedron; e.g., $\lag i,i^\pri\rag=\lag 1,2\rag$, $\lag 2,3\rag$, $\lag 3,4\rag$, and $\lag 4,1\rag$ for the $xy$ plane in \figref{fig3}. Note that there are two icosahedra in the unit cell, leading to the normalization factor $1/8$ in (\ref{6}). Supposing a perfect 90 degree order represented in \figref{fig3}(a), we obtain $O_{Cb}=1$. The cuboc order is invariant under global O(3) rotation of spin configuration. For instance, the spin configuration in \figref{fig3}(b) obtained by the MC simulation with $k_F=1.60\ \rAA^{-1}$ coincides with the perfect 90 degree order in \figref{fig3}(a) by a proper global O(3) rotation.

\begin{figure}[tbp]
 \begin{center}
  \includegraphics[width=0.5\textwidth]{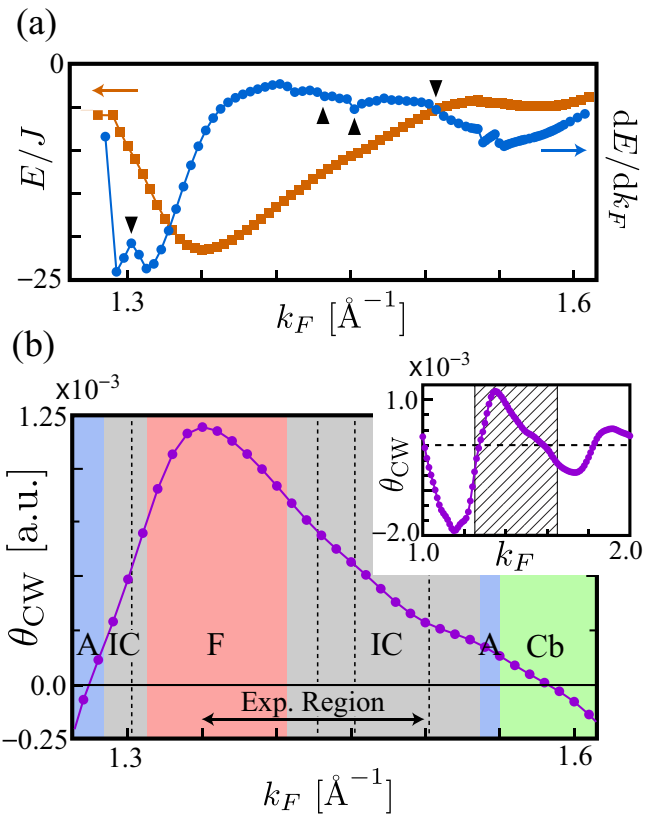}
 \end{center}
 \caption{(a) Total energy of the ground state versus the Fermi wavenumber $k_F$ and its derivative $\dr E/\dr k_F$. In the incommensurate (IC) phases, black triangles denote points where the derivative jumps/cusps indicating the the first-/second-order phase transition. Other anomalies of the derivative coincide with the phase boundaries obtained in Fig.~2.
(b) Curie-Weisse temperature $\tht_{\mrm{CW}}$ as a function of $k_F$. The inset represents $\tht_{\mrm{CW}}$ in a wider region of $k_F$, where shaded area is from $k_F=1.25~\rAA^{-1}$ to $1.65~\rAA^{-1}$. The vertical dotted lines denote the position of triangles in (a). The ferromagnetic (F), antiferromagnetic (A), cuboc (Cb), and IC phases are taken from Fig.~2. The notation `Exp. Region' indicates an region experimentally investigated~\cite{Ishikawa16,Ishikawa18}, which is estimated by electron density in the Fermi gas approximation.
 }
 \label{fig4}
\end{figure}

To confirm phase boundaries determined by the order parameters, we calculate the total energy $E$ as a function of $k_F$ and its derivative $\dr E/\dr k_F$ as shown in \figref{fig4}(a). We find clear jumps in $\dr E/\dr k_F$ at two antiferromagnetic-incommensurate boundaries with $k_F=1.29~\rAA^{-1}$ and $1.535~\rAA^{-1}$ and at a antiferromagnetic-cuboc boundary with $k_F=1.55~\rAA^{-1}$, indicating the first-order phase transition. At two incommensurate-ferromagnetic boundaries with $k_F=1.315~\rAA^{-1}$ and $1.41~\rAA^{-1}$, there is anomaly like cusp, corresponding to the second-order phase transition. In addition to the expected anomaly at the phase boundaries determined by the order parameters, we find several anomalies in $\dr E/\dr k_F$ within the incommensurate phases in \figref{fig4}(a) as denoted by black triangles. These anomalies suggest the presence of internal magnetic structures inside the incommensurate phase, which have not been detected by the order parameters examined above.  The detailed study on the internal structures remains as a future work.

In order to make clear the physical origin of magnetic phases, we plot the Curie-Weisse temperature $\tht_{\mrm{CW}}$ in \figref{fig4}(b), which is determined by the sum of total exchange energies $\tht_{\mrm{CW}}=(3N_s)^{-1}\sum_{\av{\bm{r}-\bm{r}^\pri}<R_c} J_{\av{\bm{r}-\bm{r}^\pri}}$ based on high-temperature expansion formalism. The inset in \figref{fig4}(b) shows an alternation of $\tht_{\mrm{CW}}$ between positive and negative values in the wider range of $k_F$, coming from  oscillating nature of the RKKY interaction. We find that a ferromagnetic phase appears when $\tht_{\mrm{CW}}$ is positively large as expected. On the contrary, antiferromagnetic phases emerge near $\tht_{\mrm{CW}}\cong 0$. This indicates that the antiferromagnetic state composed of ferromagnetic order inside an icosahedral cluster and antiferromagnetic spin arrangement between the neighboring clusters is achieved when the magnitude of intra-cluster ferromagnetic interactions is comparable to that of inter-cluster antiferromagnetic interactions. The cuboc phase also appears when $\tht_{\mrm{CW}}\cong 0$, indicating that ferromagnetic and antiferromagnetic interactions are totally canceled out each other but there is strong frustration as evidenced from 90 degree order. The incommensurate phase also needs strong frustration, where partial antiferromagetic interactions may play a key role.

Let us compare the calculated phases with experimentally observed ones in the Au-Al-Gd approximant. The experimental data have shown the change of phases from antiferromagnetic to ferromagnetic and to spin-glass phases with increasing $k_F$~\cite{Ishikawa16,Ishikawa18}. The region of experimentally determined $k_F$ based on the Fermi gas approximation is denoted in \figref{fig4}(b) by ``Exp. Region''.  In the region, however, our calculated phases are only ferromagnetic and incommensurate phases. Therefore, we have two inconsistencies: (i) antiferromagnetic phase observed in the experiment does not exist in the region and (ii) spin glass phase is not obtained in our calculation.

We speculate that the inconsistencies could be explained by the effect of randomness due to chemical disorders and defects in real Au-Al-Gd alloys. First, the randomness will shorten the mean free path of conduction electrons, and thus the cutoff range of the RKKY interaction, $R_c$, would be shorter than the assumed $R_c=50\ \rAA$. The shortened cutoff range suppresses absolute value of the Curie-Weisse temperature $\tht_{\mrm{CW}}$ and shifts the peak of $\tht_{\mrm{CW}}$ at $k_F=1.35\ \rAA^{-1}$ in \figref{fig4}(b) to larger $k_F$ (not shown). The change of $\tht_{\mrm{CW}}$ can narrow the ferromagnetic phase and may shift it to larger $k_F$, accompanied by the shift of antiferromagnetic phase around $k_F=1.30\ \rAA^{-1}$ possibly to the estimated experimental region. This will explain the inconsistency (i).  The randomness may also explain the inconsistency (ii) about the absence of spin glass state in our result. In general, spin glass state is induced by not only frustration but also randomness. Since the incommensurate state in our phase diagram is expected to originate from strong frustration as discussed above, including randomness in our model would bring a spin glass behavior in cooperation with strong frustration in a certain range of incommensurate phase. This is one of possible explanations of the inconsistency (ii). However, if so, we unfortunately meet another problem that experimentally observed $\tht_{\mrm{CW}}$ is negative in the spin glass phase~\cite{Ishikawa18} in contrast to the calculated positive $\tht_{\mrm{CW}}$ in our incommensurate phase. In order to resolve this new problem, we may need treat anisotropy of the crystal field around Gd ions, inhomogeneity of chemical disorders, and magnetic dipole-dipole interactions as precise as possible. This remains as an important problem to be solved in the future.

In summary, we have theoretically investigated magnetic phases in the Tsai-type quasicrystalline approximant Au-Al-Gd alloy using a minimal magnetic model including the RKKY interaction only. The change of Au/Al concentration in the alloy is assumed to correspond to the change of the Fermi wavenumber in our model. Classical MC calculations have been performed to obtain spin configurations at zero temperature. We have found a successive change among antiferromagnetic, ferromagnetic, and incommesurate phases. At large Fermi wavevector, to which real Au-Al-Gd alloy have not reached yet, we have discovered a cuboc phase. Recently such a cuboc order has attracted much attention as an exotic magnetic order experimentally observed in kagome spin compounds~\cite{Ishikawa14,Fak12} and as an origin of theoretically predicted anomalous magnetic behavior in a spin tube~\cite{Seki15,Ochiai17}. Therefore the 1/1 Tsai-type approximant Au-Al-Gd alloy is one of possible candidates for new magnetic phenomena associated with the cuboc order.

We would like to thank R. Tamura, T. J. Sato and T. Hiroto for fruitful discussions. 
This work was partly supported by Challenging Research （Exploratory） (Grant No.~JP17K18764), Grant-in-Aid for Scientific Research on Innovative Areas (Grant No.~JP19H05821). Numerical computation in this work was carried out on the Supercomputer Center at Institute for Solid State Physics, University of Tokyo and the supercomputers at JAEA.


\begin{thebibliography}{99}
\bibitem{note} The cutoff range $R_c$ should be much larger than the mean free path of conduction electrons. In Au-Al-Gd, the mean free path is estimated to be the length of lattice unit from the electric conductivity (private discussion with R. Tamura) and thus our $R_c$ satisfy the condition. In addition, $R_c=50\ \rAA$ is long enough to get convergence since the difference of total sum of exchange energy for all possible spins between $R_c=50\ \rAA$ and $60\ \rAA$ is less than $6\times 10^{-5} J$.
\end{thebibliography}
\end{document}